\documentstyle[psfig,conf_iap,10pt]{article}
\begin{document}
\heading{%
%
Modelling Galaxy Formation at high z
%
} 
\par\medskip\noindent
\author{%
Cedric Lacey$^1$, Carlton Baugh$^2$, Shaun Cole $^2$, Carlos Frenk$^2$
}
\address{%
Theoretical Astrophysics Center, Juliane Maries Vej 30, 2100
Copenhagen \OE, Denmark
}
\address{%
Department of Physics, Science Laboratories, South Rd, Durham DH1 3LE, England
}
%

\begin{abstract}
I describe a semi-analytical model for the formation and evolution of
galaxies in hierarchical clustering models, and its predictions for
the properties of the galaxy population at high z. The predictions are
found to agree well with the observed properties of the Lyman break
galaxies found at $z\sim 3$ by Steidel \et. The models
predict that the star formation rate per comoving volume should have
peaked at $z\sim 1-2$, which also agrees well with recent
observational data.
\end{abstract}
\section{Introduction }
Recent observations have begun to probe the population of normal
galaxies at high redshift. In particular, the technique of using
multi-colour surveys of faint galaxies to look for high-z objects with
strong Lyman break features has revealed a large population of
star-forming galaxies at redshifts $z\sim 3-5$, based both on
ground-based \cite{S96} and HST \cite{M96} imaging. It is a challenge
to theoretical models of galaxy formation to explain the numbers
and properties of these young galaxies. I describe here some
recent theoretical results on high-z galaxies based on semi-analytical
galaxy formation models \cite{B97}, which show that the properties of
the observed galaxies agree remarkably well with predictions based on
hierarchical clustering.

\section{Semi-analytical models}
The semi-analytical models are constructed as described in \cite{C94},
\cite{B97}. There are several steps involved. (1) We start with an
assumed comology and initial spectrum of density fluctuations. Dark
matter halos form through hierarchical clustering and merging. We
describe this by means of {\it merger trees}, constructed by a Monte
Carlo method, which specify the complete formation and merging
histories of present-day halo of different masses. The number density
of halos of different masses is given by the Press-Schechter
theory. (2) Within each halo, diffuse gas is assumed to be
shock-heated to the virial temperature during the collapse of the
halo, and then to cool radiatively. We calculate how much gas cools in
any halo during its lifetime, defined as the time until its mass has
doubled by merging, based on an assumed density profile. The gas which
cools settles into a rotationally supported disk, whose radius we
calculate based on the initial angular momentum of the gas in the
halo, derived from tidal torques. (3) The cold gas forms stars, on a
timescale which is assumed to depend on the circular velocity of the
galaxy. (4) Supernovae from massive stars reheat some of the gas,
ejecting it back into the halo, at a rate which also depends on the
circular velocity. This feedback effect strongly suppresses star
formation in low mass galaxies. The dependences of the star formation
and feedback rates on circular velocity are based on numerical
simulations \cite{NW93}. (5) In major halo mergers, when the halo mass
doubles, the remaining gas in the halo is reheated to the new virial
temperature. The largest galaxy becomes the new central galaxy, and
the galaxies from the other halos become satellite galaxies, which
then merge with the central galaxy if their dynamical friction
timescales are less than the halo lifetime. In general, not all of the
satellites will merge. Mergers between disk galaxies where the mass
ratio exceeds a critical value ($\sim 0.3$) form elliptical
galaxies. The elliptical may grow a new disk by gas cooling and become
a bulge in a spiral galaxy. (6) We calculate the luminosity evolution
of the stellar populations using an updated version of the population
synthesis model of Bruzual \& Charlot \cite{BC93}. From these models,
we calculate the distributions of galaxy masses, luminosities,
colours, sizes, circular velocities, morphologies etc, and how these
evolve with redshift.

The adjustable parameters in the model (relating to star formation,
feedback and merging) are fixed by comparison with observations of
present-day galaxies. Results of the models for zero and low redshifts
for various CDM-based cosmologies are given in \cite{C94}, \cite{H95},
\cite{B96a}, \cite{B96b}. The models fit the observed properties of
low-z galaxies (luminosity functions, colours, morphologies, number
counts) fairly well.

\section{Lyman break galaxies}
Steidel \et \cite{S96} discovered a population of normal galaxies at
$3\lsim z \lsim 3.5$, forming stars at rates $\sim 1-10 M_{\odot}/yr$,
by measuring $U_n G {\cal R}$ colours for faint galaxies and looking
for galaxies with a strong Lyman break feature at this redshift; the
redshifts were then confirmed spectroscopically. The number of such
galaxies was found to be $\approx 1400 deg^{-2}$ at ${\cal
R}_{AB}<25$. These galaxies presumably represent the progenitors of
present-day galaxies soon after they formed, and so reproducing their
properties is a crucial test of theoretical models. To perform this
test, we have used our semi-analytical models to generate mock
catalogues of galaxies covering all redshifts, including the effects
of absorption by intervening gas, and then applied the same colour
selection as Steidel \et. We present here results for 2 CDM models,
both normalized in $\sigma_8$ to match the number density of clusters
at $z=0$: an $\Omega=1$ model, with $h=0.5$, $\sigma_8=0.67$,
$\Omega_b=0.06$, and an $\Omega_0=0.3$, $\Lambda_0=0.7$ model, with
$h=0.6$, $\sigma_8=0.97$, $\Omega_b=0.04$. For a Miller-Scalo IMF,
these models predict respectively $900$ and $3000 deg^{-2}$ Lyman
break galaxies satisfying the Steidel \et colour and magnitude
selection, similar to the observed value. These values are sensitive
to the power spectrum normalization $\sigma_8$, and to the IMF
used. However, using a Salpeter IMF gives similar values. The
predicted redshift distributions match the observed ones

\begin{figure}
\centerline{\vbox{
\psfig{figure=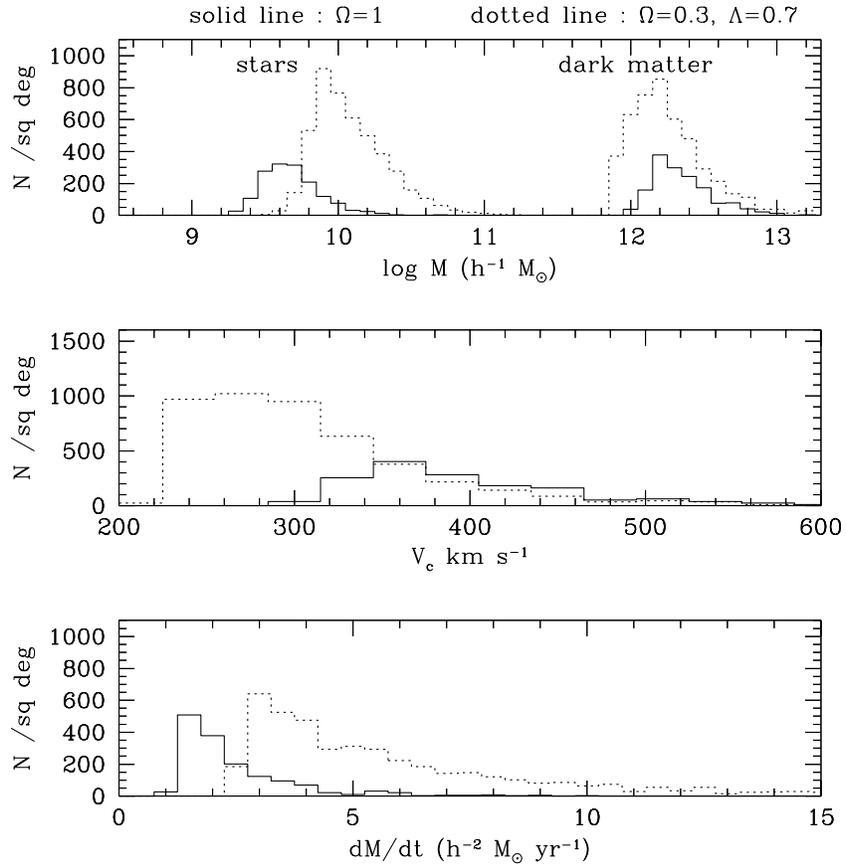,width=12.0cm}
}}
\caption[]{Predicted properties of Lyman break galaxies at $z\sim 3$,
selected according to Steidel \et $U_n G {\cal R}$ colour criteria and
with ${\cal R}_{AB}<25.0$. Results are shown for 2 CDM models,
$\Omega=1$ (solid lines) and $\Omega_0=0.3$, $\Lambda_0=0.7$ (dotted
lines). Histograms show number of objects per bin per sq.deg. Top
panel shows distributions of stellar and dark halo masses, middle
panel shows halo circular velocities, and bottom panel shows star
formation rates.}

\end{figure}

\begin{figure}
\centerline{\vbox{
\psfig{figure=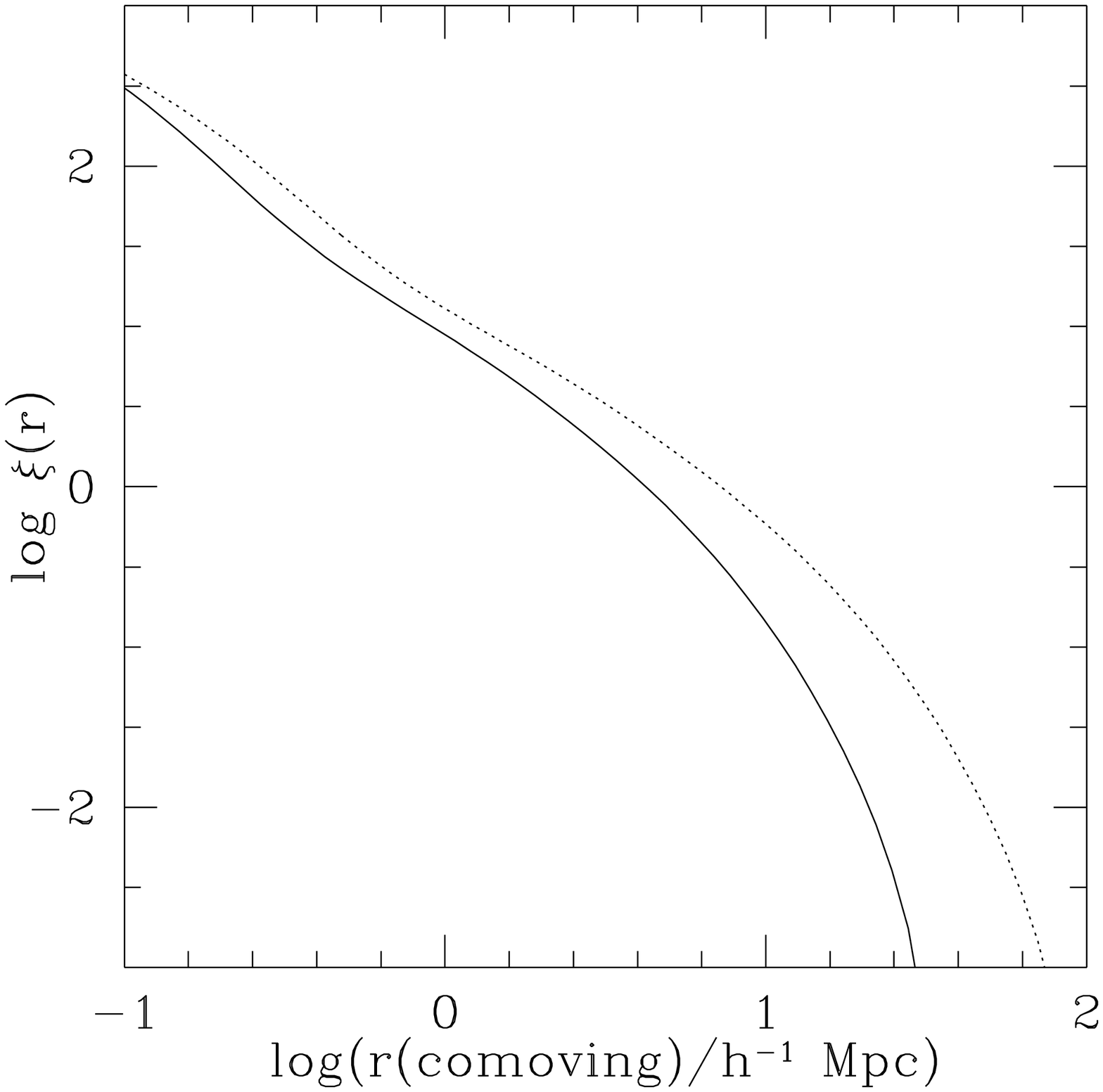,width=6.0cm}
}}
\caption[]{Predicted clustering of Lyman break galaxies at
$z\approx3$. Galaxies selected according to Steidel \et. The spatial
correlation function in comoving coordinates is shown for the
$\Omega=1$ (solid line) and $\Omega_0=0.3$, $\Lambda_0=0.7$ (dotted
line) models.  }

\end{figure}

Figure~1 shows some of the properties of Lyman break galaxies
predicted for the two CDM models. The stellar masses are predicted to
be $\sim 3\times 10^9 - 10^{10} h^{-1}M_{\odot}$, depending on the
cosmology, while the halo masses are predicted to be $\sim 10^{12}
h^{-1}M_{\odot}$ in either case. The galaxies are predicted to be in
halos with circular velocities $V_c \sim 250 - 450 km/s$, very similar
to the rotation velocities inferred observationally for these galaxies
from the widths of saturated interstellar absorption lines
\cite{S96}. The star formation rates are predicted to be $\sim 1-10
h^{-2}M_{\odot} yr^{-1}$, which follows essentially from the observed
$\cal R$ magnitudes and the assumed IMF. We also predict colours
$({\cal R} - K)_{AB}\approx 0.5-1.0$, and small sizes, with half-light
radii $\sim 0.5 h^{-1} kpc$, both of these comparable to the observed
values \cite{S96}, \cite{G96}. The halo masses predicted for these
galaxies are fairly large compared to the typical halo at $z\sim 3$,
and as a consequence, the galaxies are predicted to be quite strongly
clustered, with a bias $b\approx 4$ and a comoving correlation length
$r_0 \approx 4 h^{-1}Mpc$ (see Figure~2). This agrees with evidence
for strong clustering in the observed redshift distribution found by
\cite{S97}.

The predicted properties of the star-forming galaxy population at
$z\sim 3$ are thus in good agreement with what is known
observationally. The Lyman break galaxies found by Steidel \et are
predicted to evolve into $L\sim L_{\star}$ ellipticals and spirals at
the present day, and to be found preferentially in large groups and
clusters at $z=0$.

\section{Cosmic Star Formation History}

\begin{figure}
\centerline{\vbox{
\psfig{figure=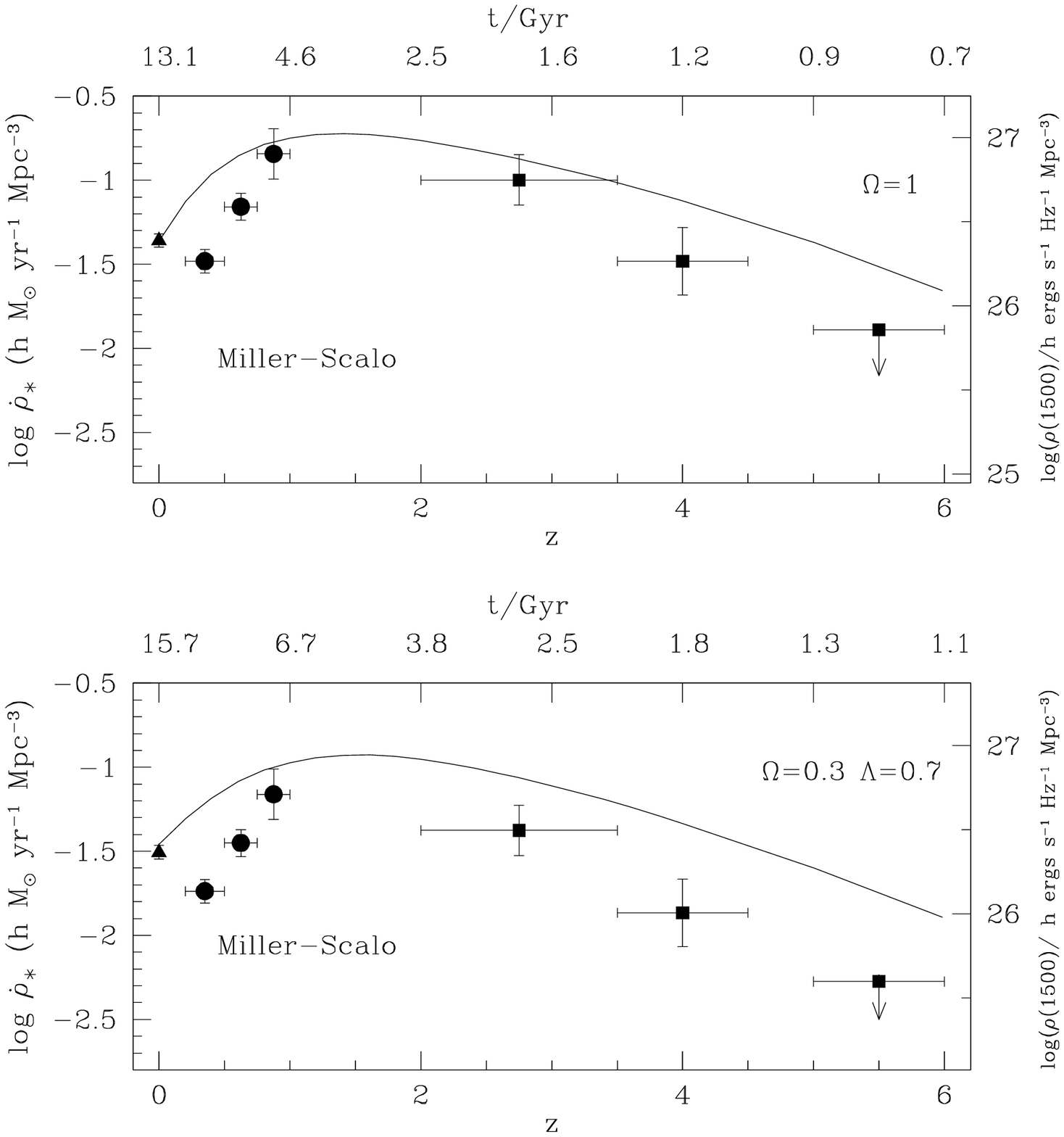,width=12.0cm}
}}
\caption[]{Cosmic star formation history. The 2 panels show the
predicted star formation rate per comoving volume vs. redshift for our
2 CDM models, compared to observational data from \cite{G95},
\cite{L96}, \cite{M96a}. The observational data have been converted to
star formation rates using the same IMF as in the models.}

\end{figure}

The global star formation history of the universe is a fundamental
quantity in galaxy formation studies. Galaxy formation models based on
CDM have typically predicted that most of the stars form fairly late,
with a median formation redshift $z<1$ according to \cite{L93},
\cite{C94}. It has recently become possible to estimate the global
star formation history out to $z\lsim 5$ observationally, using
measurements of the $H{\alpha}$ and UV luminosity functions of
galaxies at different redshifts \cite{G95}, \cite{L96},
\cite{M96a}. Significant uncertainties still remain, notably in the
form of the IMF, the correction for galaxies below the luminosity
limits of the surveys, and in the effects of dust. However, it seems
that a consistent picture is emerging. The observations indicate that
the star formation rate per comoving volume peaked at $z\sim 1-2$, and
this agrees very well with the predictions of both of our CDM models,
as shown in Figure~3. According to the models, less than 10\% of the
stars formed at $z>3$. Note that the star formation history for
$\Omega=1$ shown here is essentially identical to that predicted in an
earlier version of the model by \cite{C94}. The predictions for the
SFR history are sensitive to the parameters describing SFR and
feedback, but these were chosen as in \cite{C94} to fit observations
of present-day galaxies, without any knowledge of the observational
data in Figure~3. Our model results are therefore real predictions,
and the observational comparison represents a real success of the
model.


%
\vfill

\begin{iapbib}{99}{
\bibitem{B96a} Baugh, C.M., Cole, S., \& Frenk, C.S., 1996, \mn 282, L27
\bibitem{B96b} Baugh, C.M., Cole, S., \& Frenk, C.S., 1996, \mn 283, 1361
\bibitem{B97} Baugh, C.M., Cole, S., Frenk, C.S., \& Lacey, C.G.,
1997, \apj submitted (astro-ph/9703111)
\bibitem{BC93} Bruzual, G., \& Charlot, S., 1993, \apj 405, 538
\bibitem{C94} Cole, S., \et 1994, \mn 271, 781
\bibitem{G95} Gallego, J., Zamorano, J., Aragon-Salamanca, A., REgo,
M., 1995, \apj 455, L1
\bibitem{G96} Giavalisco, M., Steidel, S., Macchetto, F.D., 1996, \apj
470, 189.
\bibitem{H95} Heyl, J.S., Cole, S., Frenk, C.S., \& Navarro, J.F.,
1995, \mn 274, 755
\bibitem{L93} Lacey, C.G, Guiderdoni, B., Rocca-Volmerange, B., \&
Silk, J., 1993, \apj 402, 15
\bibitem{L96} Lilly, S.J., LeFevre, O., Hammer, F., \& Crampton, D.,
1996, \apj 460, L1
\bibitem{M96} Madau, P., \et 1996, \mn 283, 1388
\bibitem{M96a} Madau, P. astro-ph/9612157
\bibitem{NW93} Navarro, J.F., \& White, S.D.M., 1993, \mn 265, 271.
\bibitem{S96} Steidel, C., \et 1996, \apj 462, L17
\bibitem{S97}Steidel, C., \et 1997, \apj in press (astro-ph/9708125)
%
}
\end{iapbib}
\end{document}